\documentclass[prl,12pt]{revtex4}
\usepackage[sort&compress]{natbib}
\usepackage{graphicx}
\usepackage[FIGTOPCAP,hang,raggedright]{subfigure}
\usepackage{amsmath, amsfonts,amssymb}

\begin{document}

\title{Coherent coupling between surface plasmons and excitons in semiconductor nanocrystals}

\author{D. E. G{\'o}mez, K. C. Vernon, T. J. Davis\footnote{Corresponding author. email: tj.davis@csiro.au}}
\affiliation{CSIRO Materials Science and Engineering, Private Bag 33, Clayton South MDC, VIC, 3169 Australia}

\author{T. L. Nguyen, P. Mulvaney}
\affiliation{Bio21 Institute and School of Chemistry, The University of Melbourne, Parkville, VIC, 3010 Australia}

\begin{abstract}
We present an experimental demonstration of strong coupling between a surface plasmon propagating on a planar silver substrate, and the lowest excited state of CdSe nanocrystals. Variable-angle spectroscopic ellipsometry measurements demonstrated the formation of plasmon-exciton mixed states, characterized by a Rabi splitting of $\sim$  82 meV at room temperature. Such a coherent  interaction has the potential for the development of plasmonic non-linear devices, and furthermore, this system is akin to those studied in cavity quantum electrodynamics,  thus offering the possibility to study the regime of strong light-matter coupling in semiconductor nanocrystals at easily accessible experimental conditions. 
\end{abstract}
\maketitle


Light can be confined to the interface of a metallic surface and a dielectric, due to its interactions with the free electrons of the conductor. These evanescently-confined excitations, known as surface plasmons (SPs) \cite{maier_07,maier_AM01,maier_JoAP05,murray_AM07}, are characterised by having strong electromagnetic fields whose propagation and  energy dispersion are extremely sensitive to the dielectric properties at the metal surface.  These features, make SPs very attractive for sensing applications and for the development of nano-scale devices that can operate at very high frequencies. 

Essential elements for the practical implementation of plasmon-based technologies, such as light couplers and waveguides have been experimentally demonstrated in the past \cite{barnes_N03,engheta_S07,engheta_PRL05}. However, as in the field of electronics, the development of {\it plasmonics} will require the use of non-linear elements such as (plasmonic) diodes and transistors \cite{chang_NP07}. In this regard, the interaction of surface plasmons with optically-active materials can lead to the materialisation of such non-linear devices. 

Due to their size-dependent optical properties and, in particular, their large absorption cross-sections and high photo-stability, colloidal semiconductor nanocrystals (often referred to as quantum dots), are materials that can be used as building blocks for the development of plasmonic applications. These applications rely on the interaction of the electronic states of semiconductor nanocrystals (NCs) with the electromagnetic (EM) waves associated with SPs. 

When an SP mode is resonant with the exciton transition of an ensemble of  NCs, two regimes of interaction can be distinguished and originate due to competition between the rates of exciton-SP coupling and dephasing of either type of excitation. In the weak coupling regime, damping of either resonance dominates over coupling  and the interaction only modifies the radiative decay rate of the exciton state (Purcell effect) and its  angular pattern of radiation.
Many reports exist in the literature that illustrate this coupling regime for the case of semiconductor NCs. 
In the weak--coupling regime, the SP--exciton interaction has been shown to result in enhancements of the excitation rate of the NCs \cite{chen_APL08}, and of their fluorescence intensity \cite{song_NL05,pompa_NN06,soganci_OE07,matsuda_APL08,yeh_APL08,zhang_PRB05} an effect even observable at  the single NC level \cite{shimizu_prl02,ito_PRB07}.
Furthermore, the photoluminescence of semiconductor NCs has been employed for the excitation of propagating SPs in metallic nanostructure \cite{akimov_N07,fedutik_PRL07,fedutik_JotACS07}, and due to the large absorption cross-section of the NCs, it has  also been possible to attenuate SP propagation in a structure containing these nanoscale light absorbers \cite{pacifici_NP07}.

On the contrary, in the strong coupling regime, the states of the SP--exciton system are coherent superpositions of both types of excitations and are expected to have significantly different optical properties. For instance, in the case of  a single exciton transition, these mixed states are characterized by  a doublet of SP-exciton ``polaritons", that should be readily observable in transmittance and reflectivity measurements. Although this SP--exciton coupling regime has been reported for silicon nanocrystals \cite{takeda_JoAP07}, dye molecules \cite{pockrand_JoCP82}, organic semiconductors (such as in J-aggregates) \cite{bellessa_PRL04,dintinger_PRB05,sugawara_PRL06} and quantum well structures \cite{weisbuch_PRL92, symonds_APL07, vasa_PRL08,bellessa_PRB08},  direct experimental evidence of coherent coupling of excitons in colloidal semiconductor NCs with SPs has not been documented. 

In this letter, we present an experimental demonstration of such coherent energy coupling between  SPs on a thin Ag film and excitons confined in semiconductor NCs. This coupling is evident in the energy dispersion of the coupled system due to the appearance of  the anti-crossing of two ``polariton-like" branches, as derived from variable-angle, spectroscopic ellipsometry measurements.

\begin{figure}[ht!]
\centering
\includegraphics[width=.85\textwidth]{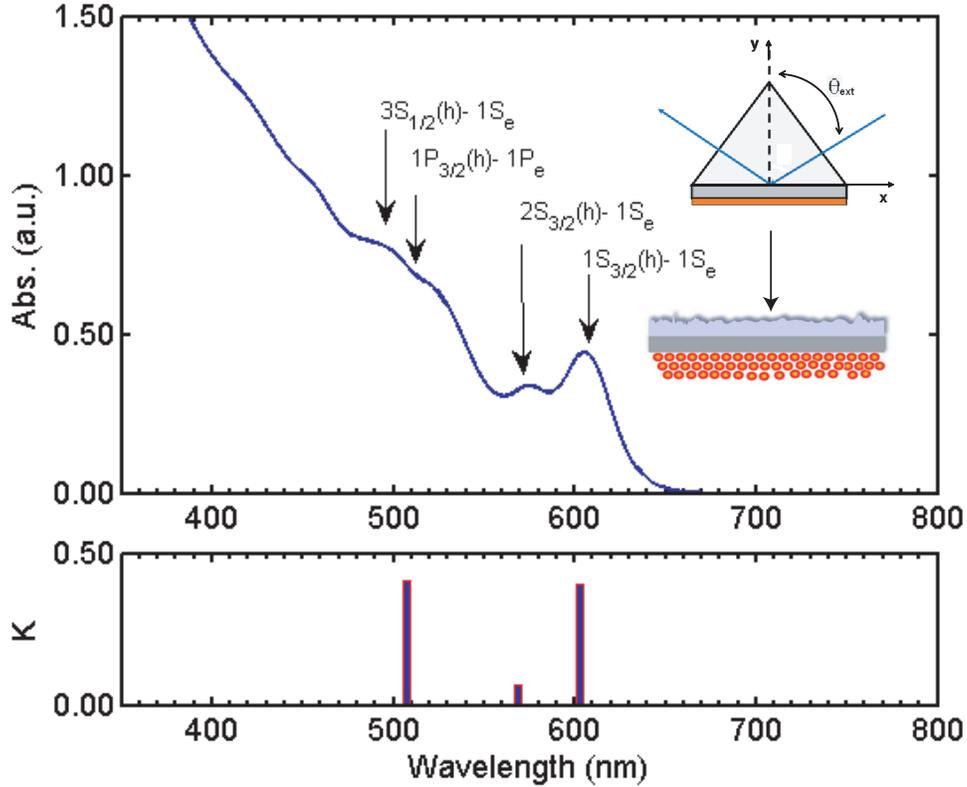}
\caption{Normal incidence absorption spectrum of CdSe films spun cast onto glass slides. The vertical lines in the bottom panel, show the positions of the exciton transitions and their e-h overlap (relative to that of the $1S_{3/2}(h)1S_e$, defined in eqn. \eqref{eq:K}) , calculated by: (i) taking into account the valence band degeneracy, ignoring the effects of coupling to the split-off band, (ii) a finite confining potential for the electron and (iii)  the electron-hole Coulomb interaction \cite{ekimov_josab93}. The states are labelled as $\mathrm{n_hL_F\,n_eL_e}$, where $\mathrm{L_e}$ is the angular momentum of the electron envelope function, whereas $\mathrm{L_F}$ is the minimum {\it orbital} momentum of the hole envelope wavefunction with {\it total} momentum $\mathrm{F = L_h + J}$ ($\mathrm J$ is the unit--cell angular momentum). The average diameter of the NCs is 4.8 nm as estimated by using the calibration curve of ref. \cite{yu_CoM03}. The inset shows a schematic diagram of the thin films and of  the prism-coupling geometry, where $\theta$ denotes the angle of incidence.} 
\label{fig:1}
\end{figure}

Thin films consisting of Ag and SiO$_2$ were deposited onto glass coverslips by e-beam deposition of each material at a base pressure of $\precsim$ 2$\times$10$^{-6}$ Torr, to achieve a nominal thickness of 50 nm (Ag) and 3nm (SiO$_2$) (as monitored with a quartz crystal oscillator).  The SiO$_2$ layer was added onto the Ag film in order to prevent its degradation. CdSe NCs were synthesised in 1-octadecene (ODE) following standard literature methods \cite{embden_l05}. After careful removal of the ODE (by ligand exchange with 5-amino-1-pentanol \cite{jasieniak_afm07}), the NCs were dispersed in isopropanol and filtered through a 0.45 $\mu$m filter. Small quantities of this stock solution were subsequently spin-coated directly onto the Glass/Ag/SiO$_2$ films at a speed of 2000 rpm to achieve a typical film thickness of  74 nm (measured with a surface profilometer, mean surface roughness 5.63 nm). An optical absorption spectrum of such a CdSe film on  bare glass is shown in figure \ref{fig:1}. 

The glass slides containing the Glass/Ag/SiO$_2$/CdSe films were attached to a right-angle prism (BK7, Thorlabs) with immersion oil (through the glass side of the film, see inset of figure \ref{fig:1}) and mounted on the sample stage of a photo-elastically-modulated, variable-angle, spectroscopic ellipsometer (Uvisel, JY-Horiba). In this ellipsometer, the angle of incidence could be varied from 55$^\circ$ to 90$^\circ$ (1$^\circ$ increments, measured from the normal to the sample stage). Illumination was provided by a 100 W Xe-arc lamp.

In ellipsometry, the ratio ($\rho$) of the complex Fresnel reflection coefficient for p- and s-polarised light is measured as a function of the incident wavelength ($\lambda$) for a given range of angles of incidence ($\theta$). This quantity $\rho$, is defined as:
\begin{equation}
    \rho \equiv \frac{r_\text{p}}{r_\text{s}} = \biggl |\frac{r_\text{p}}{r_\text{s}}\biggr | e^{i(\Delta_\text{p}-\Delta_\text{s})} = \tan(\Psi)\,e^{i\Delta},
\label{eq:1}
\end{equation}
\noindent
where $\Psi$ and $\Delta$ are the ellipsometric angles   which determine the changes of the magnitude and phase of an incident beam of light, upon reflection from an optical system \cite{azzam_ellips}. 

\begin{figure}[ht!]
\centering
\subfigure[]{\includegraphics[width=.5\textwidth]{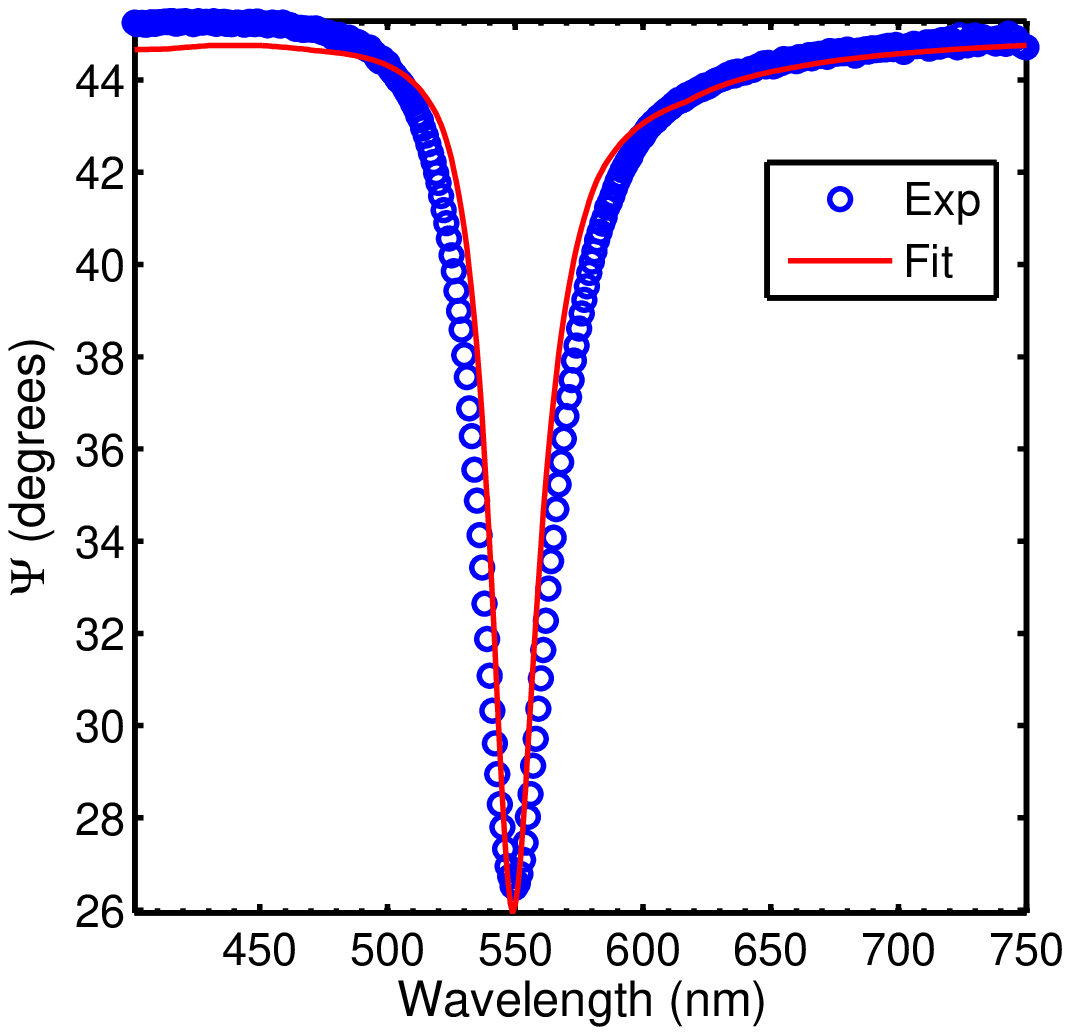}}
\subfigure[]{\includegraphics[width=.5\textwidth]{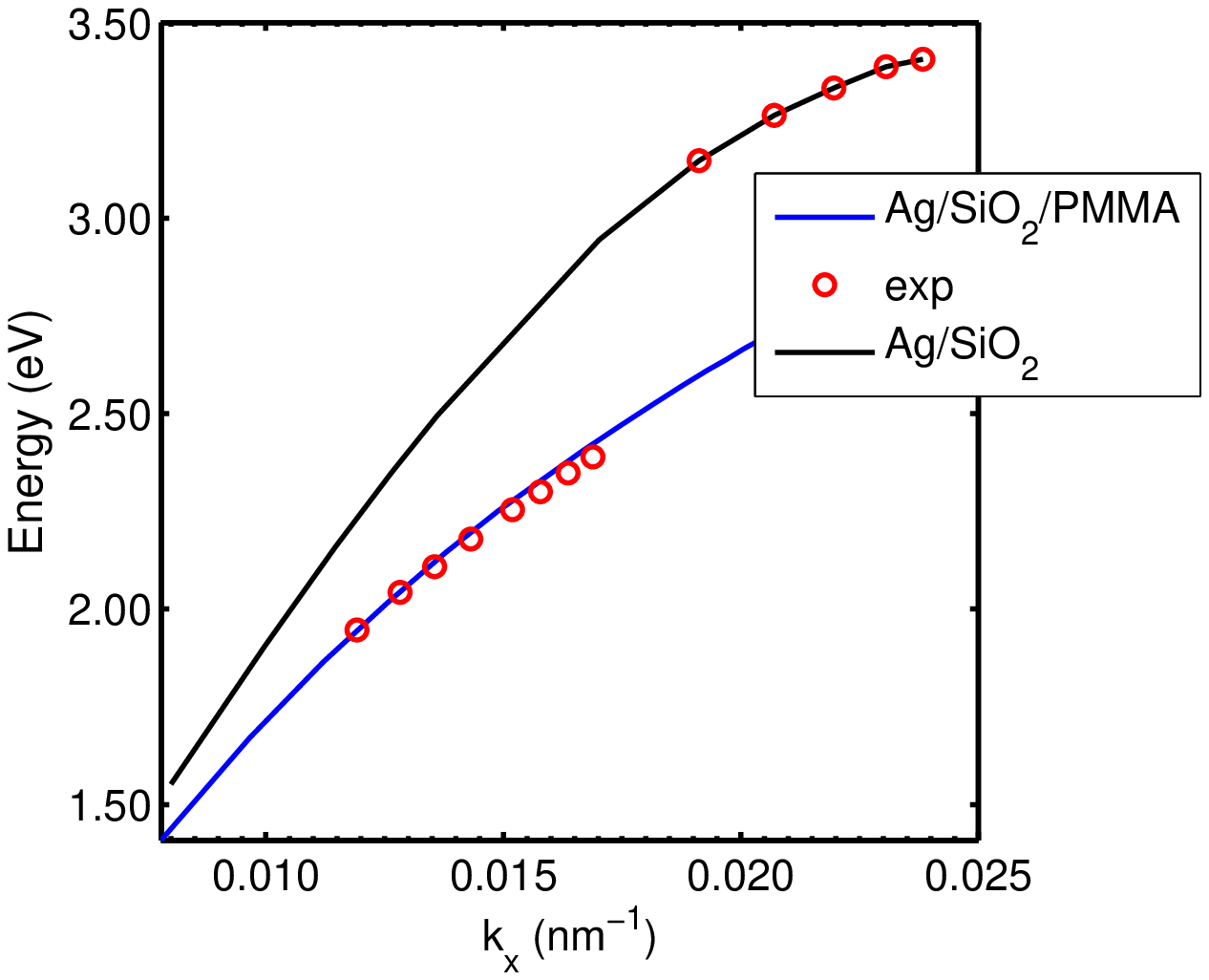}}
\caption{(a) Spectrum of $\Psi$ [absolute value of $\rho$, eqn. \eqref{eq:1}], measured via ellipsometry under the Kretschmann configuration, for thin films of Ag/SiO$_2$/poly(methyl methacrylate) (PMMA). The red line is a fit to the data with the tranfer-matrix method \cite{born_opticsBook} (derived thicknesses from the fit are 49 nm, 4 nm and 51.6 nm for Ag/SiO$_2$/PMMA respectively). The angle of incidence, with respect to the sample normal (external to the prism) was 70$^\circ$. (b) Energy dispersion of the SP in Ag/SiO$_2$ and in Ag/SiO$_2$/PMMA film.  The lines were calculated using the transfer-matrix method The blue line corresponds to the energy dispersion obtained when the films are coated with 51.6 nm of PMMA. In this figure, the in-plane wavevector was calculated as $k_x = (2\pi/\lambda)\sin(\theta) n_G$, with $n_G$ the refractive index of the prism (BK7 $n$ = 1.515 at 633 nm).}
\label{fig:2}
\end{figure}

The measurement of  $\Psi$ for a film consisting of Glass/Ag/SiO$_2$/PMMA is shown in figure \ref{fig:2}(a)  where a dip can be noticed at a wavelength of $\approx$ 550 nm. This dip corresponds to a minimum in the reflectivity of p-polarised light that arises from the excitation of a SP at the surface of the Ag film. Such an excitation can be described by Maxwell's equations (with the transfer matrix method \cite{born_opticsBook}). Using literature values for the dielectric data of Ag \cite{johnson_PRB72}, SiO$_2$ \cite{palik_85} and PMMA (measured to be 1.485) leads to the fit shown as the red line of figure 2(a). 

The wavelength position of the SP resonance can be tuned by changing the angle of incidence or equivalently, the in-plane wavevector ($k_x$) of the incident light beam, as exemplified in figure 2(b).
In this figure we present the SP energy dispersion for both a bare Glass/Ag/SiO$_2$ and a Glass/Ag/SiO$_2$/PMMA film, demonstrating in this way, not only the tunability of the SP energy  by changes in $k_x$, but also by the interaction of the SP with a thin layer of a dielectric material. In both cases the energy of the SP increases with $k_x$, but it is readily seen that at any value of $k_x$, the energy of the SP in the Ag/SiO$_2$/PMMA film is smaller than the one found for the Ag/SiO$_2$ case.

\begin{figure}[ht!]
\centering
\includegraphics[scale=.80]{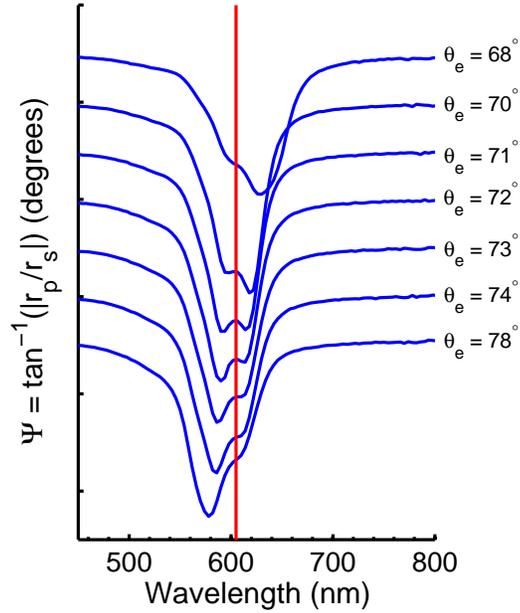}
\vspace{.5cm}
\caption{Reflectivity spectrum, shown as $\Psi$ defined in eqn. \eqref{eq:1},  for a film consisting of Ag/SiO$_2$/CdSe measured at different angles of incidence as indicated. The vertical red line shows the position of the lowest excited state of the CdSe NCs. The reflectivity data has been shifted vertically for clarity.}
\label{fig:reflectivity_NCs}
\end{figure}

When the Ag/SiO$_2$ film is coated with CdSe NCs, the reflectivity spectrum  shows two dips, that as can be seen on figure \ref{fig:reflectivity_NCs}, change both in shape and wavelength position with the angle of incidence. Initially, at angles around and below (not shown) 60$^\circ$ the reflectivity is dominated by one dip at wavelengths above the red line, but at 71$^\circ$, the two reflectivity minima have almost equal contributions and are energetically separated by $\sim$ 82.14 meV. As the angle of incidence is increased beyond this point, the contribution from the highest--energy dip to the spectra increases, and the wavelength positions of the minima in the two dips show an avoided crossing at the wavelength indicated with the red line.


The first important observation to be made in connection to figure \ref{fig:reflectivity_NCs}, is that the reflectivity spectra consists of only two features. This implies, that although there are several exciton levels in each CdSe NC in the film (see figure 1),  the SPs only couple to one. From the position of the avoided crossing in figure \ref{fig:reflectivity_NCs}, it is evident that this exciton level is the lowest excited state of the NCs, namely the $1S_{3/2}(h)1S_e$ exciton state. 

\begin{figure}[ht!]
\centering
\includegraphics[scale=.65]{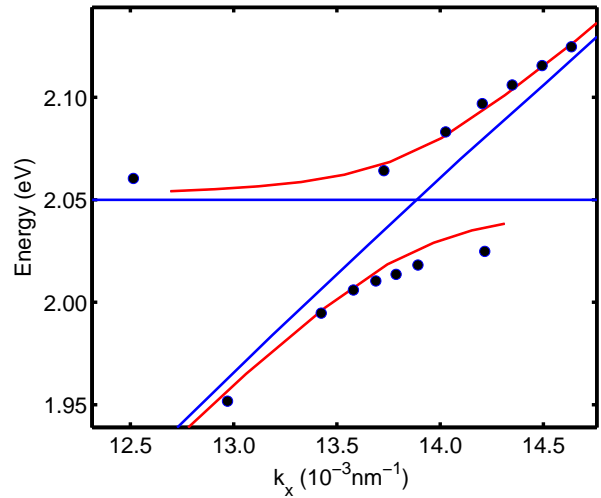}
\caption{Energy dispersion curve,  derived from the measured  dips in $\Psi$. The horizontal axis is the in-plane wavevector component $k_x$ for a Ag/SiO$_2$/CdSe film. The dots correspond to the experimental data. The blue lines correspond to the dispersionless exciton energy (2.05 eV) and the un--coupled SP dispersion (see supplementary information). The red lines are the branches $E_U$ and $E_L$ predicted by equation \eqref{eq:coupled_osc} by setting $\gamma_X = \gamma_{SP} = 0$ and with a Rabi splitting of $\sim$ 80 meV, without adjusting any of these parameters.}
\label{fig:3}
\end{figure}

In figure \ref{fig:3}, we present the energy position of the dips in the reflectivity spectra of figure \ref{fig:reflectivity_NCs} as a function of the in-plane wavevector $k_x$. The data in figure \ref{fig:3}, clearly shows the appearance of two branches that anti-cross in $k_x$, an unambiguous demonstration of {\it coherent coupling} between the SP supported by the Ag/SiO$_2$ film, and the exciton state in the CdSe film. These branches have a minimum energy gap at about 0.014 nm$^{-1}$ and anti-cross at the energy of the $1S_{3/2}(h)1S_e$ exciton  of the CdSe NCs (approximately 2.05 eV). Away from this resonance, each branch on the dispersion curve approaches asymptotically the decoupled exciton and SP modes respectively.

Within the picture of coupled oscillators, the energy of the exciton-SP system  at any value of wavevector $k_x$  is given by \cite{andreani_PRB99,reithmaier_N04}:
\begin{equation}\label{eq:coupled_osc}
E_{(U,L)}(k_x) = \frac{E_{SP}(k_x) + E_X}2 - i\hbar \frac{\gamma_X+\gamma_{SP}}4\pm \hbar \bigg\{g^2 +\frac 1 4\bigg(E_{SP}(k_x) - E_X + i\frac{\gamma_{SP}-\gamma_X}2\bigg )^2\bigg\}^{1/2},
\end{equation}
\noindent where $E_{SP}(k_x)$ is the energy of the un--coupled SPs,  $E_X$ that of the (dispersionless) exciton states, $\hbar\gamma_{SP}$ and $\hbar\gamma_X$ are the  SP and exciton spectral full-width at half-maximum (FWHM) and $g$ is a measure of  the interaction strength between the two energy modes. 
This constant is given by the matrix element of the operator ${\bf d\cdot E}$, evaluated between the initial and final state of the system, $g = |\langle{\bf d\cdot E}\rangle|/\hbar$, where $\bf{d}$ is the excited state dipole moment of the NCs,  and $\bf{E}$ the magnitude, at the position of the NCs, of the electric field due to the SPs.  

Strong exciton-SP coupling occurs when the  coupling strength $g$ dominates over damping of either type of  excitation, or stated otherwise, when the coupling constant $g$ satisfies the following condition: $g^2 > (\gamma_{SP} - \gamma_X)^2/16$. In this case, the energy dispersion curve consists of two branches ($E_U$ and $E_L$), as observed in figure \ref{fig:3}, that are separated by a (vacuum) Rabi splitting given by $2\hbar\sqrt{g^2 - (\gamma_{SP} - \gamma_X)^2/16}$, which we found to be $\approx$ 80 meV.

Damping of the SP resonance in the Ag film  occurs mainly by ohmic losses in the metal and by radiative losses that arise from imperfections at the surface of the film. For bare Glass/Ag/SiO$_2$ and Glass/Ag/SiO$_2$/PMMA films the spectral FWHM was found to be  $\hbar\gamma_{SP} = 150$ meV, which according to our ongoing discussion, implies that the following condition must hold in order to observe strong coupling: $\hbar g > |150 \text{ meV} - \hbar\gamma_X|/4$. 

Within the effective mass approximation for the exciton states in semiconductor NCs and the  electric--dipole approximation for their interaction with the SPs, the magnitude of  $g$ is proportional to the oscillator strength of the exciton transition, which in turn is proportional to the electron-hole wavefunction overlap $K$ \cite{ekimov_josab93}:
\begin{equation}\label{eq:K}
    K = \bigg |\int dr\,r^2\,f_e(r)\,f_h(r)\bigg |^2, 
\end{equation}
\noindent where $f_{e,h}(r)$ are the radial components of the electron (e) and hole (h) envelope functions. Due to spherical symmetry and the lack of nodes in $f_{e,h}(r)$ inside the NC volume, this overlap integral $K$ has its maximum value for the $1S_{3/2}(h)1S_e$ exciton state (see figure 1), partially accounting  for the observation of strong SP coupling to this state. Higher excited states can also have a non-negligible value of the overlap integral, but as stated before,  the observation of avoided crossings in the energy dispersion of the coupled system must result from  a balance between this coupling strength $g$ (and therefore the e-h overlap) and the  FWHM of the exciton transition $\gamma_X$ (i.e. when $\hbar g > |150 \text{ meV} - \hbar\gamma_X|/4$). It has been found experimentally that  exciton transition linewidths increase with increasing energy \cite{norris_PRB96}, which may translate into stronger exciton damping and therefore into a violation of the condition for the observation of strong coupling to these states. Furthermore, it is known that excitation of the NC into higher exciton states results in ultra-fast non--radiative relaxation to the lowest excited state \cite{klimov_prb99} which, in light of the experiments discussed here, implies that these states are strongly damped and therefore should not exhibit strong coupling to the SPs.

Up to this point, the discussion applies to the interaction between a single NC and the SP resonance. In the NC films there are (at least) two types of disorder that can affect the coupling strength $g$: (i) the inherent size distribution that results from the NC synthesis and (ii)  the disorder that arises from fluctuations in the concentration of NCs in the  film.

Due to the quantum size effect, the energy of the exciton states in NCs vary with the size of these structures. Colloidal NCs have an intrinsic size and shape distribution that is expected to introduce inhomogeneous broadening of the exciton transitions. However, with current literature methods the synthesis of semiconductor NCs  results typically in a dispersion no larger than 5\% in the average size. Spin-coating of the CdSe NCs onto the films is expected to result in a disordered layer where the inter-particle separations are not perfectly homogeneous. If the distance between the NCs is small enough, direct energy transfer between these could compete with coupling to the SP \cite{kim_PRB08}, possibly providing a source of damping. Furthermore, an inhomogeneity in the concentration of NCs could lead to the formation of large aggregates that could act as scattering centres for the propagation of SPs. 
Our observation of a finite Rabi splitting indicates  that in the NC film these parameters play a minor role in the dynamics of the coupled system, and therefore in the NC films there is low inter-particle coupling, which would otherwise make the system extremely sensitive to the spatial variation on the film. On the contrary, the existence of such finite Rabi splitting indicates that a large fraction of the NCs in the film interacts {\it in phase} with the SP electric field. 

In the past, experiments aimed at demonstrating strong light-matter interactions with semiconductor NCs have been carried out by resonantly coupling their photoluminescence to a single mode of a high-Q micro-cavity \cite{thomas_NL06}. However, due to the limited {\it emission} oscillator strength of CdSe NCs \cite{leistikow_PRB09}, the reported Rabi splitting was smaller (30 - 45 $\mu$eV) than the one found in the present study. This difference arises from the fact that, although the effective cavity Q value for SPs is low ($\sim$ 10), the electric field strength at the position of the NC film is relatively high, this being a consequence of the strong spatial localization of SPs at the surface of the metal film. Furthermore, in the system presented here,  coupling occurs to the first absorbing state of the NCs which carries most of the absorption oscillator strength.


In summary, we have presented an experimental demonstration of  strong coupling between SPs in a thin Ag film and excitons in a film of semiconductor NCs. The observation of a Rabi splitting in this system implies that a significant fraction of the NCs in the film coherently exchanges energy with the electromagnetic mode of the metal. Remarkably, these oscillations were observed  at room temperature and under low excitation powers, making this system attractive for future applications, in particular, for the development of exciton-plasmon non-linear devices and  low threshold lasers. 

\begin{acknowledgments}

We would like to express our gratitude to B. Sexton for his help with the physical vapour deposition of the Ag films.
\end{acknowledgments}


\begin{thebibliography}{10}
\expandafter\ifx\csname url\endcsname\relax
  \def\url#1{{\tt #1}}\fi
\expandafter\ifx\csname urlprefix\endcsname\relax\def\urlprefix{URL }\fi
\providecommand{\eprint}[2][]{\url{#2}}

\bibitem{maier_07}
S.~Maier.
\newblock {\em Plasmonics: fundamentals and applications\/} (Springer, New
  York, 2007).

\bibitem{maier_AM01}
S.~A. Maier, M.~L. Brongersma, P.~G. Kik, S.~Meltzer, A.~A.~G. Requicha and
  H.~A. Atwater.
\newblock Advanced Materials {\bf 13}, 1501 (2001).

\bibitem{maier_JoAP05}
S.~A. Maier and H.~A. Atwater.
\newblock Journal of Applied Physics {\bf 98}, 011101 (pages~10) (2005).

\bibitem{murray_AM07}
W.~Murray and W.~Barnes.
\newblock Advanced Materials {\bf 19}, 3771 (2007).

\bibitem{barnes_N03}
W.~L. Barnes, A.~Dereux and T.~W. Ebbesen.
\newblock Nature {\bf 424}, 824 (2003).

\bibitem{engheta_S07}
N.~Engheta.
\newblock Science {\bf 317}, 1698 (2007).

\bibitem{engheta_PRL05}
N.~Engheta, A.~Salandrino and A.~Al{\`u}.
\newblock Physical Review Letters {\bf 95}, 95504 (2005).

\bibitem{chang_NP07}
D.~E. Chang, A.~S. Sorensen, E.~A. Demler and M.~D. Lukin.
\newblock Nature Physics {\bf 3}, 807 (2007).

\bibitem{chen_APL08}
Y.~Chen, K.~Munechika, I.~J.-L. Plante, A.~M. Munro, S.~E. Skrabalak, Y.~Xia
  and D.~S. Ginger.
\newblock Applied Physics Letters {\bf 93}, 053106 (pages~3) (2008).

\bibitem{song_NL05}
J.-H. Song, T.~Atay, S.~Shi, H.~Urabe and A.~Nurmikko.
\newblock Nano Letters {\bf 5}, 1557 (2005).

\bibitem{pompa_NN06}
P.~P. Pompa, L.~Martiradonna, A.~D. Torre, F.~D. Sala, L.~Manna, M.~D.
  Vittorio, F.~Calabi, R.~Cingolani and R.~Rinaldi.
\newblock Nat Nano {\bf 1}, 126 (2006).

\bibitem{soganci_OE07}
I.~M. Soganci, S.~Nizamoglu, E.~Mutlugun, O.~Akin and H.~V. Demir.
\newblock Opt. Express {\bf 15}, 14289 (2007).

\bibitem{matsuda_APL08}
K.~Matsuda, Y.~Ito and Y.~Kanemitsu.
\newblock Applied Physics Letters {\bf 92}, 211911 (pages~3) (2008).

\bibitem{yeh_APL08}
D.-M. Yeh, C.-F. Huang, Y.-C. Lu and C.~C. Yang.
\newblock Applied Physics Letters {\bf 92}, 091112 (pages~3) (2008).

\bibitem{zhang_PRB05}
J.~Zhang, Y.-H. Ye, X.~Wang, P.~Rochon and M.~Xiao.
\newblock Physical Review B {\bf 72}, 201306 (pages~4) (2005).

\bibitem{shimizu_prl02}
K.~T. Shimizu, W.~K. Woo, B.~R. Fisher, H.~J. Eisler and M.~G. Bawendi.
\newblock Physical Review Letters {\bf 89}, 117401 (2002).

\bibitem{ito_PRB07}
Y.~Ito, K.~Matsuda and Y.~Kanemitsu.
\newblock Physical Review B {\bf 75}, 033309 (pages~4) (2007).

\bibitem{akimov_N07}
A.~V. Akimov, A.~Mukherjee, C.~L. Yu, D.~E. Chang, A.~S. Zibrov, P.~R. Hemmer,
  H.~Park and M.~D. Lukin.
\newblock Nature {\bf 450}, 402 (2007).

\bibitem{fedutik_PRL07}
Y.~Fedutik, V.~V. Temnov, O.~Schops, U.~Woggon and M.~V. Artemyev.
\newblock Physical Review Letters {\bf 99}, 136802 (pages~4) (2007).

\bibitem{fedutik_JotACS07}
Y.~Fedutik, V.~Temnov, U.~Woggon, E.~Ustinovich and M.~Artemyev.
\newblock Journal of the American Chemical Society {\bf 129}, 14939 (2007).

\bibitem{pacifici_NP07}
D.~Pacifici, H.~J. Lezec and H.~A. Atwater.
\newblock Nature Photonics {\bf 1}, 402 (2007).

\bibitem{takeda_JoAP07}
E.~Takeda, M.~Fujii, T.~Nakamura, Y.~Mochizuki and S.~Hayashi.
\newblock Journal of Applied Physics {\bf 102}, 023506 (pages~6) (2007).

\bibitem{pockrand_JoCP82}
I.~Pockrand, A.~Brillante and D.~M\"{o}bius.
\newblock J Chem. Phys. {\bf 77}, 6289 (1982).

\bibitem{bellessa_PRL04}
J.~Bellessa, C.~Bonnand, J.~C. Plenet and J.~Mugnier.
\newblock Phys. Rev. Lett. {\bf 93}, 036404 (2004).

\bibitem{dintinger_PRB05}
J.~Dintinger, S.~Klein, F.~Bustos, W.~L. Barnes and T.~W. Ebbesen.
\newblock Physical Review B {\bf 71}, 035424 (pages~5) (2005).

\bibitem{sugawara_PRL06}
Y.~Sugawara, T.~A. Kelf, J.~J. Baumberg, M.~E. Abdelsalam and P.~N. Bartlett.
\newblock Physical Review Letters {\bf 97}, 266808 (pages~4) (2006).

\bibitem{weisbuch_PRL92}
C.~Weisbuch, M.~Nishioka, A.~Ishikawa and Y.~Arakawa.
\newblock Phys. Rev. Lett. {\bf 69}, 3314 (1992).

\bibitem{symonds_APL07}
C.~Symonds, J.~Bellessa, J.~C. Plenet, A.~Br\'{e}hier, R.~Parashkov, J.~S.
  Lauret and E.~Deleporte.
\newblock Applied Physics Letters {\bf 90}, 091107 (pages~3) (2007).

\bibitem{vasa_PRL08}
P.~Vasa, R.~Pomraenke, S.~Schwieger, Y.~I. Mazur, V.~Kunets, P.~Srinivasan,
  E.~Johnson, J.~E. Kihm, D.~S. Kim, E.~Runge, G.~Salamo and C.~Lienau.
\newblock Physical Review Letters {\bf 101}, 116801 (pages~4) (2008).

\bibitem{bellessa_PRB08}
J.~Bellessa, C.~Symonds, C.~Meynaud, J.~C. Plenet, E.~Cambril, A.~Miard,
  L.~Ferlazzo and A.~Lema\^{i}tre.
\newblock Physical Review B {\bf 78}, 205326 (2008).

\bibitem{embden_l05}
J.~van Embden and P.~Mulvaney.
\newblock Langmuir {\bf 21}, 10226 (2005).

\bibitem{jasieniak_afm07}
J.~Jasieniak, J.~Pacifico, R.~Signorini, F.~Maurizio, A.~Martucci and
  P.~Mulvaney.
\newblock Adv. Funct. Mater {\bf 17}, 1654 (2007).

\bibitem{azzam_ellips}
R.~Azzam and N.~Bashara.
\newblock {\em {Ellipsometry and polarized light}\/} (North-Holland, 1977).

\bibitem{born_opticsBook}
M.~Born and E.~Wolf.
\newblock {\em {Principles of optics}\/}.
\newblock 3 ed. (Pergamon Press New York, 1964).

\bibitem{johnson_PRB72}
P.~B. Johnson and R.~W. Christy.
\newblock Phys. Rev. B {\bf 6}, 4370 (1972).

\bibitem{palik_85}
E.~Palik, ed.
\newblock {\em {Handbook of Optical Constants of Solids}\/} (Academic Press,
  1985).

\bibitem{andreani_PRB99}
L.~C. Andreani, G.~Panzarini and J.-M. G\'erard.
\newblock Phys. Rev. B {\bf 60}, 13276 (1999).

\bibitem{reithmaier_N04}
J.~P. Reithmaier, G.~Sek, A.~Loffler, C.~Hofmann, S.~Kuhn, S.~Reitzenstein,
  L.~V. Keldysh, V.~D. Kulakovskii, T.~L. Reinecke and A.~Forchel.
\newblock Nature {\bf 432}, 197 (2004).

\bibitem{ekimov_josab93}
A.~I. Ekimov, F.~Hache, M.~C. Schanne-Klein, D.~Ricard, C.~Flytzanis, I.~A.
  Kudryavtsev, T.~V. Yazeva, A.~V. Rodina and A.~L. Efros.
\newblock Journal of the Optical Society of America B: Optical Physics {\bf
  10}, 100 (1993).

\bibitem{norris_PRB96}
D.~J. Norris and M.~G. Bawendi.
\newblock Phys. Rev. B {\bf 53}, 16338 (1996).

\bibitem{klimov_prb99}
V.~I. Klimov, D.~W. McBranch, C.~A. Leatherdale and M.~G. Bawendi.
\newblock Physical Review B {\bf 60}, 13740 (1999).

\bibitem{kim_PRB08}
D.~Kim, S.~Okahara, M.~Nakayama and Y.~Shim.
\newblock Physical Review B {\bf 78}, 153301 (pages~4) (2008).

\bibitem{thomas_NL06}
N.~L. Thomas, U.~Woggon, O.~Sch\"ops, M.~V. Artemyev, M.~Kazes and U.~Banin.
\newblock Nano Letters {\bf 6}, 557 (2006).

\bibitem{leistikow_PRB09}
M.~D. Leistikow, J.~Johansen, A.~J. Kettelarij, P.~Lodahl and W.~L. Vos.
\newblock Physical Review B {\bf 79}, 045301 (pages~9) (2009).

\bibitem{yu_CoM03}
W.~W. Yu, L.~Qu, W.~Guo and X.~Peng.
\newblock Chemistry of Materials {\bf 15}, 2854 (2003).


\end{thebibliography}

\section{Figures}

\end{document}